\documentclass[aps,prd,reprint,superscriptaddress,nofootinbib]{revtex4-1}
\usepackage{amssymb,amsmath,graphicx,hyperref}
\usepackage[utf8]{inputenc}

\def\be{\begin{equation}}
\def\ee{\end{equation}}
\def\bea{\begin{eqnarray}}
\def\eea{\end{eqnarray}}
\def\beal{\begin{equation}\begin{aligned}}
\def\eeal{\end{aligned}\end{equation}}
\def\bse{\begin{subequations}}
\def\ese{\end{subequations}}
\def\nn{\nonumber}

\def\bra#1{\langle #1|}
\def\ket#1{|#1 \rangle}
\def\braket#1{\langle #1 \rangle}
\def\la{\lambda}
\def\lb{\tilde{\lambda}}

\def\bs{\boldsymbol}
\def\Res_#1{\operatorname*{Res}_{#1}}
\def\sgn{\operatorname*{sgn}}

\def\ie{i.e. }
\def\eg{e.g. }

\def\eqn#1{Eq.~\eqref{#1}}
\def\eqns#1#2{Eqs.~\eqref{#1} and~\eqref{#2}}

\def\Eqn#1{Eq.~\eqref{#1}}

\def\fig#1{Fig.~{\ref{#1}}}

\def\rcite#1{Ref.~\cite{#1}}
\def\rcites#1{Refs.~\cite{#1}}

\begin{document}

\title{Black-hole scattering with general spin directions
\\ from minimal-coupling amplitudes}

\author{Alfredo Guevara}
\email[]{aguevara@pitp.ca}
\affiliation{Perimeter Institute for Theoretical Physics, Waterloo, ON N2L 2Y5, Canada}
\affiliation{Department of Physics \& Astronomy, University of Waterloo, Waterloo, ON N2L 3G1, Canada}
\affiliation{CECs Valdivia \& Departamento de F\'isica, Universidad de Concepci\'on, Casilla 160-C, Concepci\'on, Chile}

\author{Alexander Ochirov}
\email[]{aochirov@phys.ethz.ch}
\affiliation{ETH Z\"urich, Institut f\"ur Theoretische Physik,
Wolfgang-Pauli-Str. 27, 8093 Z\"urich, Switzerland}

\author{Justin Vines}
\email[]{justin.vines@aei.mpg.de}
\affiliation{Max Planck Institute for Gravitational Physics (Albert Einstein Institute), Am M\"uhlenberg 1, Potsdam 14476, Germany}

\date{\today}

\begin{abstract}
We study the link between classical scattering of spinning black holes and quantum amplitudes for massive spin-$s$ particles. Generic spin orientations of the black holes are considered, allowing their spins to be deflected on par with their momenta. We rederive the spin-exponentiated structure of the relevant tree-level amplitude from minimal coupling to Einstein's gravity, which in the $s\to\infty$ limit generates the black holes' complete series of spin-induced multipoles. The resulting scattering function is seen to encode in a simple way the known net changes in the black-hole momenta and spins at first post-Minkowskian order. We connect our findings to a rigorous framework developed elsewhere for computing such observables from amplitudes.
\end{abstract}

\maketitle

\section{Introduction
\label{sec:intro}}

Applying techniques from quantum field theory to the study of
the classical two-body problem in general relativity (GR) has produced
significant progress in the treatment of the post-Newtonian (PN) approximation
\cite{Holstein:2004dn,Goldberger:2004jt,Porto:2005ac,Levi:2015msa,
Porto:2016pyg,Levi:2018nxp}, expanding in small speeds and weak fields.
A particular focus of more recent interest has been
the use of quantum scattering amplitudes to produce explicit classical results 
in the post-Minkowskian (PM) approximation~\cite{Neill:2013wsa,Damour:2016gwp},
which resums the expansion in small speeds while still expanding in weak coupling.
Several new contributions have been made to the understanding of
conservative monopolar two-body dynamics up to 2PM order
(through ${\cal O}(G^2)$, where $G$ is Newton's constant)
using one-loop amplitudes~\cite{Cachazo:2017jef,Guevara:2017csg,Damour:2017zjx,
Bjerrum-Bohr:2018xdl,Cheung:2018wkq,KoemansCollado:2019ggb,Cristofoli:2019neg},
all ultimately confirming an equivalent classical solution
from decades ago~\cite{Westpfahl:1985}.
A significant milestone has been the first presentation of results at 3PM order
from a two-loop amplitude calculation~\cite{Bern:2019nnu},
using an arsenal of modern amplitude techniques
\cite{Bern:1994zx,Bern:1994cg,Britto:2004nc,Kawai:1985xq,Bern:2008qj,Johansson:2014zca}.

The first results for spin-orbit coupling in the PM scheme
have been computed only recently, from purely classical considerations,
at 1PM and 2PM orders~\cite{Bini:2017xzy,Bini:2018ywr}.
Spin-orbit (or dipole) effects are universal in the two-body problem in GR,
in that their form does not depend on the nature of the spinning bodies.
It was shown in \rcites{Bini:2017xzy,Bini:2018ywr} how they are determined
by the parallel transport map along the geodesic worldline
in the (regularized) spacetime metric sourced by monopolar bodies.

Going beyond the pole-dipole level,
higher-multipole couplings specific to black holes (BHs)
were treated at 1PM order in \rcite{Vines:2017hyw},
by means of a classical effective action approach~\cite{Levi:2015msa,Porto:2016pyg}
matched to the linearized Kerr solution,
to all orders in the spin-induced multipole expansion.
In this paper we aim to fully reproduce the central results of \rcite{Vines:2017hyw}
from amplitudes, exploiting and further substantiating the remarkable fact
that the BH multipole structure up to the $2^{2s}$-pole level is reproduced
by considering spin-$s$ particles which are minimally coupled to gravity.
This was first demonstrated in \rcite{Vaidya:2014kza} up to the spin-2 level
for the leading-PN-order corrections to the two-body interaction potential,
following work along similar lines in \rcites{Holstein:2008sw,Holstein:2008sx}.

A generalization of minimal-coupling amplitudes to arbitrary spins~$s$
has been proposed recently in \rcite{Arkani-Hamed:2017jhn}
using a new massive spinor-helicity formalism.
Through the course of
\rcites{Guevara:2017csg,Guevara:2018wpp,Chung:2018kqs,Bautista:2019tdr},
there has emerged a consistent picture of how these amplitudes encode
the complete tower of spinning-BH multipole moments, at least at 1PM order,
when one lets the spin quantum number~$s$ tend to infinity.\footnote{We note that \rcites{Guevara:2017csg,Guevara:2018wpp,Chung:2018kqs} have also treated spin contributions at 2PM order, and that \rcite{Bautista:2019tdr} has also considered radiative effects via a classical double copy with spin.}
Concerning explicit specifications of the classical two-body dynamics,
the final results of \rcite{Chung:2018kqs} allowed arbitrary spin orientations
but restricted to leading orders in the PN expansion,
while those of \rcite{Guevara:2018wpp} were complete at 1PM order
but restricted to the case where the BH spin vectors
are aligned with the orbital angular momentum.
In this aligned-spin case, the two-body scattering is confined
to a single constant orbital plane, and the spin vectors are conserved,
pointing orthogonally to that plane.
In the general case, both the orbital plane and the spin vectors
are rotated in the course of the interaction.  

In this paper we consider general spin directions at full 1PM order.
The staring point is the tree-level amplitude for one-graviton exchange
between two massive spin-$s$ particles shown in \fig{fig:4pt}.
We compute its holomorphic classical limit (HCL) \cite{Guevara:2017csg}
by gluing two of the minimal-coupling three-point amplitudes~\cite{Arkani-Hamed:2017jhn}
depicted in \fig{fig:3pt}.
We streamline the treatment of the spin-exponentiated structure of such amplitudes,
which was the focus of \rcite{Guevara:2018wpp},
by incorporating additional Lorentz boosts~\cite{Bautista:2019tdr}
into the spin exponentials.
Finally, we adapt to our needs a general formalism~\cite{Kosower:2018adc,Maybee:2019jus}
for extracting gauge-invariant classical observables from amplitudes.
It has already been used in \rcite{Maybee:2019jus}
to compute the net changes in the momenta and spins
for two-body scattering at 1PM order,
reproducing the results for BHs up to quadrupolar order from minimally coupled spin-1 particles.
Here we extend such calculation to arbitrary spins $s$, and in the limit $s\to\infty$ obtain all orders
in the BH multipole expansions at 1PM order \cite{Vines:2017hyw}.

\section{Minimal coupling to gravity
\label{sec:review}}

In this section we review the angular-momentum exponentiation
that is inherent to the gravitational coupling of spinning black holes
and the corresponding amplitudes.

At the linearized-gravity level, the classical stress-energy tensor serving as an effective source for a single Kerr black hole with mass $m$, classical momentum $p^\mu = m u^\mu$
and spin $S^\mu = m a^\mu$ is~\cite{Vines:2017hyw}
\be
   T^{\mu\nu}_\text{BH}(x) = \frac{1}{m}\!\int\!d\tau\;\!
      p^{(\mu} \exp(a*\partial)^{\nu)}_{~~\rho} p^\rho \delta^{4}(x-u \tau) ,
\label{eq:KerrEnergyTensor}
\ee
where we have used the shorthand notation
\be   
   (a*b)^{\mu\nu} = \epsilon^{\mu\nu\alpha\beta} a_\alpha b_\beta .
\label{eq:LeviCivitaStar}
\ee
The spin transversality condition $p\cdot a=0$ is also assumed.
The corresponding coupling of the BH to gravity is
\beal
\label{eq:KerrEnergyTensorCoupling}
   S_\text{BH} &
    = -\frac{\kappa}{2} \int\!\hat{d}^4k\;\!h_{\mu\nu}(k)
      T^{\mu\nu}_\text{BH}(-k) \\ &
    = -\kappa\!\int\!\hat{d}^4k\;\!\hat{\delta}(2p \cdot k)
      p^{(\mu} \exp(a*ik)^{\nu)}_{~~\rho} p^\rho h_{\mu\nu}(k) ,
\eeal
where the coupling constant is $\kappa=\sqrt{32 \pi G}$.
Here and below the hats over the delta functions and measures encode
appropriate positive or negative powers of $2\pi$, respectively.
Putting the graviton on shell,
$h_{\mu\nu}(k) \to \hat{\delta}(k^2) \varepsilon_\mu \varepsilon_\nu$,
we can rewrite the characteristic angular-momentum exponential
in another form
\beal\!
   h_{\mu\nu}(k) T^{\mu\nu}_\text{BH}(-k)
    = \hat{\delta}(k^2) \hat{\delta}(p \cdot k) (p\cdot\varepsilon)^2
      \exp\!\bigg(\!{-i}\frac{k_\mu \varepsilon_\nu S^{\mu\nu}}
                             {p\cdot\varepsilon}\bigg) ,
\label{eq:KerrEnergyTensorExp}
\eeal
now involving a transverse spin tensor $S^{\mu\nu}$
\be
\label{eq:SpinTensorSSC}
   S^{\mu\nu}=\epsilon^{\mu\nu\rho\sigma} p_\rho a_\sigma
   \qquad \Rightarrow \qquad
   p_\mu S^{\mu\nu} = 0 .
\ee
More explicitly, the above transition relies on the equality
\be
\label{eq:KerrEnergyTensorExpDerivation}
   (p\cdot\varepsilon)^{j-1} \varepsilon_\mu \big[(a*ik)^j\big]^\mu_{~\,\nu} p^\nu
    = \big({-i}k_\mu \varepsilon_\nu S^{\mu\nu}\big)^j ,
\ee
that is easiest verified in the frame and gauge where
$k=(k^0,0,0,k^0)$,
$\varepsilon=(0,\varepsilon^1,\pm i\varepsilon^1,0)$ and
$p=(p^0,p^1,0,p^0)$.

\begin{figure}
\includegraphics[width=0.245\textwidth]{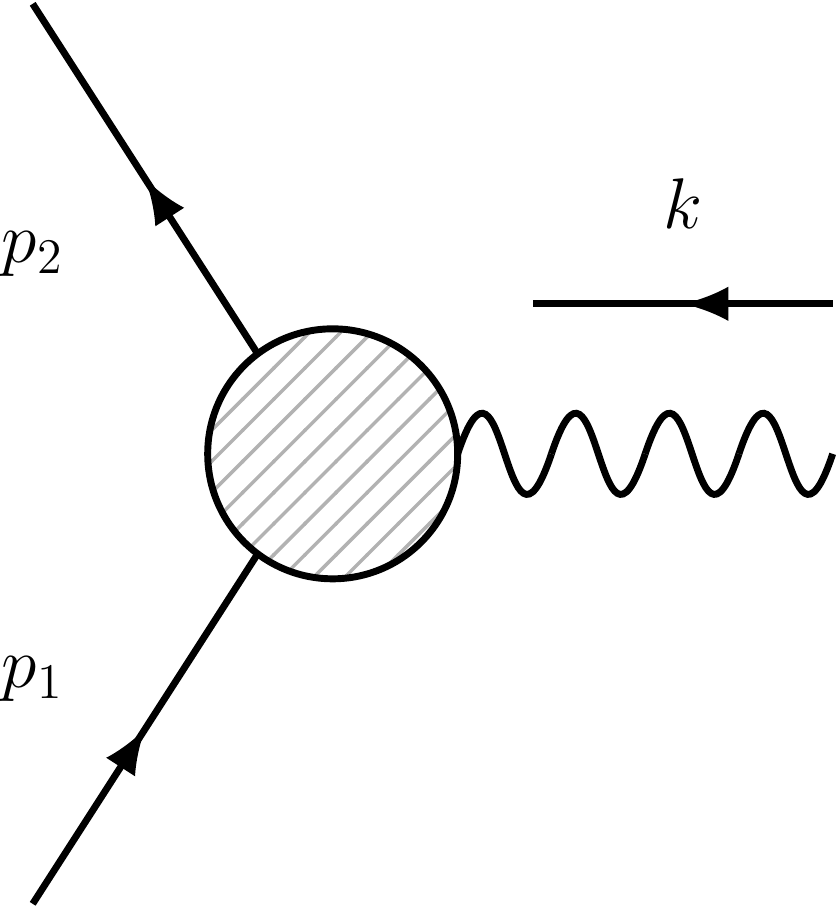}
\caption{\label{fig:3pt} Three-point amplitude}
\end{figure}

In \rcite{Guevara:2018wpp} we discovered
that the same exponential is hidden inside
the minimal-coupling amplitudes
proposed recently by Arkani-Hamed, Huang and Huang~\cite{Arkani-Hamed:2017jhn}
\bse
\begin{align}
\label{eq:GravityMatter3ptPlus}
   {\cal M}_3(p_1^{\{a\}}\!,-p_2^{\{b\}}\!,k^{+}) & = -\frac{\kappa}{2}
      \frac{\braket{1^a 2^b}^{\odot 2s}}{m^{2s-2}} x^2 , \\*
\label{eq:GravityMatter3ptMinus}
   {\cal M}_3(p_1^{\{a\}}\!,-p_2^{\{b\}}\!,k^{-}) & = -\frac{\kappa}{2}
      \frac{[1^a 2^b]^{\odot 2s}}{m^{2s-2}} x^{-2} .
\end{align} \label{eq:GravityMatter3pt}%
\ese
The arguments of scattering amplitudes are treated as incoming,
so the present choice corresponds to the momentum configuration
shown in \fig{fig:3pt}.
Here and below the symbol $\odot$ denotes tensor product symmetrized
over each massive particle's $2s$ little-group indices
$\{a_1,\ldots,a_{2s}\}$ and $\{b_1,\ldots,b_{2s}\}$.
Furthermore, $x$ is the positive-helicity factor
\be
   x = \frac{[k|p_1\ket{r}}{m \braket{k\;\!r}}
     = -\frac{\sqrt{2}}{m} (p_1\cdot\varepsilon^+)
     = \bigg[ \frac{\sqrt{2}}{m} (p_1\cdot\varepsilon^-) \bigg]^{-1} ,
\label{eq:xFactor}
\ee
that is dimensionless and independent of the reference momentum~$r$
on the on-shell three-point kinematics~\cite{Arkani-Hamed:2017jhn}.

Now one can start by noticing
that the amplitudes ${\cal M}_3^{(0)}=-\kappa (p_1\cdot\varepsilon^\pm)^2$,
given by the scalar case of \eqn{eq:GravityMatter3pt},
correspond precisely to the $S^{\mu\nu}=0$ case
of the vertex~\eqref{eq:KerrEnergyTensorExp}.
Moreover, we were able to
recast the spin structure of the amplitudes~\eqref{eq:GravityMatter3pt}
in an exponential form~\cite{Guevara:2018wpp}
\bse
\begin{align}
\label{eq:GravityMatter3ptExpPlus}
   \braket{2^b 1^a}^{\odot 2s} &
    = [2^b|^{\odot 2s}\!
      \exp\!\bigg(\!{-i}\frac{k_\mu \varepsilon_\nu^+ \bar{\sigma}^{\mu\nu}}
                        {p_1\cdot\varepsilon^+}\bigg) |1^a]^{\odot 2s} , \\
\label{eq:GravityMatter3ptExpMinus}
   [2^b 1^a]^{\odot 2s} &
    = \bra{2^b}^{\odot 2s}\!
      \exp\!\bigg(\!{-i}\frac{k_\mu \varepsilon_\nu^- \sigma^{\mu\nu}}
                        {p_1\cdot\varepsilon^-}\bigg) \ket{1^a}^{\odot 2s} ,
\end{align} \label{eq:GravityMatter3ptExp}%
\ese
featuring a tensor-product version of the
chiral and antichiral spinorial generators
\be
   \sigma^{\mu\nu} = \frac{i}{4}
      [\sigma^\mu \bar{\sigma}^\nu\!- \sigma^\nu \bar{\sigma}^\mu] , \qquad
   \bar{\sigma}^{\mu\nu} = \frac{i}{4}
      [\bar{\sigma}^\mu \sigma^\nu\!- \bar{\sigma}^\nu \sigma^\mu] .
\label{eq:LorentzGeneratorSpinor}
\ee
In \rcite{Guevara:2018wpp}
we could translate between the spin-operator exponentials~\eqref{eq:GravityMatter3ptExp}
and the classical-spin exponential~\eqref{eq:KerrEnergyTensorExp}
and thus identify the minimal-coupling amplitudes~\eqref{eq:GravityMatter3pt}
with Kerr black holes.
(A complementary identification was done in \rcite{Chung:2018kqs}
by matching to the Wilson coefficients
in the one-body effective field theory of a Kerr black hole~\cite{Levi:2015msa,Levi:2018nxp}.)
Such a translation involved sticking to either chiral or antichiral representation,
so that one of the amplitudes in \eqn{eq:GravityMatter3pt} contains
no apparent dependence on the spin operator.
The correct symmetric dependence on the classical spins was restored
in \rcite{Guevara:2018wpp} via a notion of generalized expectation value (GEV),
which involved division by the product of the polarization tensors
of the incoming and outgoing avatars of the BH.
In this paper we follow an alternative path:
we recover the entire spin information from the amplitudes
by combining the spinor-helicity formalism
with the covariant approach to multipoles introduced in \rcite{Bautista:2019tdr},
while the GEV will only serve to fix the normalization.

\section{General integer-spin setup
\label{sec:spin}}

Although the minimal-coupling amplitudes~\eqref{eq:GravityMatter3pt}
of \rcite{Arkani-Hamed:2017jhn} are valid for both integer and half-integer spins,
for simplicity we will concentrate on the former case.
Spin-$s$ polarization tensors are constructed
as~\cite{Guevara:2018wpp,Chung:2018kqs}
\be
   \varepsilon_{p\:\!\mu_1\ldots\mu_s}^{a_1\ldots a_{2s}}
    = \varepsilon_{p\:\!\mu_1}^{(a_1 a_2} \ldots
      \varepsilon_{p\:\!\mu_s}^{a_{2s-1} a_{2s})} , \qquad
   \varepsilon_{p\:\!\mu}^{ab}
    = \frac{i\bra{p^{(a}}\sigma_\mu|p^{b)}]}{\sqrt{2}m} .
\label{eq:PolTensors}
\ee
We adopt the spinor-helicity conventions of
\rcite{Ochirov:2018uyq},
so the spin-1 polarization vectors are spacelike
and obey the standard properties that are expected from them:
\bse
\begin{align}
   p\cdot\varepsilon_{p}^{ab} & = 0 , \\
\label{eq:PolTensorsCompleteness}
   \varepsilon_{p\:\!\mu}^{ab} \varepsilon_{p\:\!\nu ab} &
    =-\eta_{\mu\nu} + \frac{p_\mu p_\nu}{m^2} , \\
\label{eq:PolTensorsNorm}
   \varepsilon_{p\:\!11}\cdot\varepsilon_{p}^{11} &
    = \varepsilon_{p\:\!22}\cdot\varepsilon_{p}^{22}
    = 2\:\!\varepsilon_{p\:\!12}\!\cdot\!\varepsilon_{p}^{12} = -1 , \\
\label{eq:PolTensorsConjugation}
   (\varepsilon_{p\:\!\mu}^{ab})^* & = \varepsilon_{p\:\!\mu\:\!ab}
    = \epsilon_{ac} \epsilon_{bd} \varepsilon_{p\:\!\mu}^{cd} .
\end{align}
\label{eq:PolTensorsProperties}%
\ese
In particular, the last line follows from the conjugation rule
\be
   (\la_{p\:\!\alpha}^{~\;a})^*
    = \sgn(p^0) \lb_{p\:\!\dot{\alpha}\:\!a}
   \quad \Leftrightarrow \quad
   (\lb_{p\:\!\dot{\alpha}}^{~\;a})^*
    =-\sgn(p^0) \la_{p\:\!\alpha\:\!a} ,
\label{eq:MassiveSpinorConjugation}
\ee
which implements the fact that in the little group ${\rm SU}(2)$
upper and lower indices are related by complex conjugation.

Since the polarization tensors are essentially
symmetrized tensor products $\varepsilon_p^{\odot s}$,
the action of the Lorentz generators
is trivially induced by the vector representation
\be
   \Sigma^{\mu\nu,\sigma}_{~~~~~\tau}
    = i[\eta^{\mu\sigma} \delta^\nu_\tau - \eta^{\nu\sigma} \delta^\mu_\tau] ,
\label{eq:LorentzGenerator1}
\ee
namely,
\beal
   (\Sigma^{\mu\nu})^{\sigma_1\ldots\sigma_{s}}_{~~~~~~~\,\tau_1\ldots\tau_{s}}
    = \Sigma^{\mu\nu,\sigma_1}_{~~~~~\,\tau_1}
      \delta^{\sigma_2}_{\tau_2}\!\ldots \delta^{\sigma_s}_{\tau_s} & \\ + \ldots
    + \delta^{\sigma_1}_{\tau_1}\!\ldots \delta^{\sigma_{s-1}}_{\tau_{s-1}}
      \Sigma^{\mu\nu,\sigma_s}_{~~~~~\,\tau_s} & .
\label{eq:LorentzGenerator}
\eeal
These matrices realize the Lorentz algebra on the one-particle states of spin~$s$,
which are represented by the polarization tensors~\eqref{eq:PolTensors}.

A more convenient spin quantity to deal with is the Pauli-Lubanski vector
\be
   S_{\lambda} =
      \frac{1}{2m} \epsilon_{\lambda\mu\nu\rho} S^{\mu\nu} p^{\rho} .
\label{eq:PauliLubanski}
\ee
Here $S^{\mu\nu}$ is the spin tensor,
the transverse part of which can be reconstructed from the vector as
\be
   S_\perp^{\mu\nu} = \frac{1}{m} \epsilon^{\mu\nu\rho\sigma} p_\rho S_\sigma
   \quad \Rightarrow \quad
   p_\mu S_\perp^{\mu\nu} = 0 .
\label{eq:PauliLubanski2}
\ee
Understanding \eqn{eq:PauliLubanski} in the operator sense,
we can derive the general form of one-particle matrix element
of the Pauli-Lubanski spin operator,
here denoted by $\Sigma^{\mu}$,
\begin{align}
\label{eq:PauliLubanskiLGOperator}
 & \varepsilon_p^{\{a\}}\!\cdot \Sigma^\mu\!\cdot \varepsilon_p^{\{b\}}
    = \frac{s(-1)^{s-1}}{2m} \\~& \times\!
      \big\{ \bra{p^{(a_1}}\sigma^\mu|p^{(b_1}]
           + [p^{(a_1}|\bar{\sigma}^\mu\ket{p^{(b_1}} \big\}
      \epsilon^{a_2 b_2}\!\ldots \epsilon^{a_{2s}) b_{2s})} . \nn
\end{align}
One way to give meaning to this formula
is to lower one set of indices and set it equal to the other:
it then produces an expectation value
\be
\label{eq:PauliLubanskiLGEV}
   \frac{ \varepsilon_{p\{a\}}\!\cdot \Sigma^\mu\!\cdot \varepsilon_p^{\{a\}} }
        { \varepsilon_{p\{a\}}\!\cdot \varepsilon_p^{\{a\}} }
    =\!\left\{\!
      \begin{aligned}
      s & s_p^\mu ,~~a_1 = \ldots = a_{2s} = 1 , \\
      (s-1) & s_p^\mu ,~~{\textstyle\sum_{j=1}^{2s}} a_j = 2s+1 , \\
      (s-2) & s_p^\mu ,~~{\textstyle\sum_{j=1}^{2s}} a_j = 2s+2 , \\
      \dots \\
      -s & s_p^\mu ,~~a_1 = \ldots = a_{2s} = 2 ,
      \end{aligned}
      \right.\!\!
\ee
where we have also accounted for the nontrivial normalization of the tensors.
This shows that spin is quantized
in terms of the unit-spin vector
\be
   s_p^\mu = -\frac{1}{2m}
      \big\{ \bra{p_1}\sigma^\mu|p^1]
           + [p_1|\bar{\sigma}^\mu\ket{p^1} \big\}
\label{eq:UnitSpinVector}
\ee
that is transverse and spacelike, $p \cdot s_p = 0$, $s_p^2 = -1$.
This vector is familiar from textbook discussions of the Dirac spin,
in which context it may be written as
$ \frac{1}{2m} \bar{u}_{p\:\!1} \gamma^\mu \gamma^5 u_p^1
=-\frac{1}{2m} \bar{u}_{p\:\!2} \gamma^\mu \gamma^5 u_p^2 $.
According to \eqn{eq:PauliLubanskiLGEV},
this vector corresponds to the spin quantization axis
and identifies the $(2s+1)$ distinct wavefunctions
$\varepsilon_p^{1 \ldots 1 2 \ldots 2}$ with states of definite spin projection.

Moving on towards the scattering context,
let us consider a three-point kinematics $p_1+k=p_2$ shown in \fig{fig:3pt}.
A naive extension of the spin matrix element~\eqref{eq:PauliLubanskiLGOperator},
now between states with different momenta of mass~$m$, is
\begin{align}
\label{eq:SpinTransitionNaive}\!\!\!
   \varepsilon_2^{\{b\}}\!\cdot \Sigma^\mu[p_{\rm a}] \cdot \varepsilon_1^{\{a\}}
    = &\;\frac{s}{4m^{2s}}
      \big\{ \bra{1^a}\sigma^\mu|2^b]
           + [1^a|\bar{\sigma}^\mu\ket{2^b} \big\} \\ \odot\,
      \big\{ \braket{1^a 2^b}\,- &\,[1^a 2^b] \big\} \odot
      \braket{1^a 2^b}^{\odot(s-1)}\!\odot [1^a 2^b]^{\odot(s-1)} . \nn
\end{align}
Here we have encoded the symmetrization of the little-group indices
into the modified tensor-product symbol~$\odot$,
and the indices on the right-hand side should be regarded as abstract placeholders.
It is important to stress that the symmetrization encoded in the symbol $\odot$
only acts inside the two ${\rm SU}(2)$-index sets $\{a\}$ and $\{b\}$ separately,
as symmetrizing a little-group index of momentum $p_1$
with that of $p_2$ would be mathematically inconsistent.

Notice that in \eqn{eq:SpinTransitionNaive} we must specify that
the Pauli-Lubanski operator
is defined with respect to the average momentum $p_{\rm a}=(p_1+p_2)/2$.
It is this momentum that we will associate with the classical momentum
$p_{\rm a}^\mu = m u_{\rm a}^\mu$ of one of the incoming black holes,
so it makes sense to define a spin vector to be orthogonal to it.
\Eqn{eq:SpinTransitionNaive}
treats the chiral and antichiral spinors on an equal footing
and generalizes the spin-1 matrix element considered in \rcite{Guevara:2018wpp}.
However, even in the spin-1 case
the angular-momentum exponentiation~\eqref{eq:GravityMatter3ptExp},
present in the exclusively chiral and antichiral spinorial representations,
was found to be opaque at the level of such a matrix element.
The reason for that is physically important.
As discussed in \rcite{Levi:2015msa},
a consistent picture of spin-induced multipoles of a pointlike particle
must be formulated in the particle's rest frame,
in which the spin does not precess~\cite{Weinberg:1972kfs}.
Therefore, the formula~\eqref{eq:SpinTransitionNaive} is too naive,
as it involves a spin operator defined for momentum $p_{\rm a}$
but acts with it on the states with momenta $p_{\rm a} \pm k/2$.
The cure for that is to take into account
additional Lorentz boosts, which we will now proceed to do.

\section{Angular-momentum exponentiation
\label{sec:spinexponents}}

Bautista and one of the current authors
have recently argued that all spin multipoles of the amplitude
can be extracted through a finite Lorentz boost~\cite{Bautista:2019tdr}.
This boost is needed to bridge the gap between two states
with different momenta.
In this way, the quantum picture is made
consistent with the classical notion of spin-induced multipoles
of a pointlike object on a worldline~\cite{Levi:2015msa}.
Here we introduce such a construction
in terms of the spinor-helicity variables.
Its equivalence to the covariant formalism of \rcite{Bautista:2019tdr} is explained in the Appendix.

To start, we note that any two four-vectors $p_1$ and $p_2$ of equal mass $m$
may be related by 
\be
   p_2^\rho = \exp\!\big( i \mu_{12} p_1^\mu p_2^\nu \Sigma_{\mu\nu}
                    \big)^\rho_{~\sigma} p_1^\sigma ,
\label{eq:Lorentz1to2vector}
\ee
where we have used the generators~\eqref{eq:LorentzGenerator1}. The numeric prefactor in the exponent is explicitly
\be
   \mu_{12}
    = \frac{ \log\!\big[ \frac{1}{m^2}
                         \big(p_1 \cdot p_2 {+} \sqrt{(p_1 \cdot p_2)^2 {-} m^2}\big)
                   \big] }
           { \sqrt{(p_1 \cdot p_2)^2 - m^2} }
 = \frac{1}{m^2} {+} \mathcal{O}(k^2) .
\ee
Here we are only interested in the strictly on-shell setup,
for which $k^2=(p_2-p_1)^2=0$.
The corresponding spinorial transformations are
\bse
\begin{align}
\label{eq:Lorentz1to2Spa}
   \ket{2^b} & = U_{12~a}^{~~b}
      \exp\!\bigg( \frac{i}{m^2} p_1^\mu k^\nu \sigma_{\mu\nu}\!\bigg)
      \ket{1^a} , \\
\label{eq:Lorentz1to2Spb}
   |2^b] & = U_{12~a}^{~~b}
      \exp\!\bigg( \frac{i}{m^2} p_1^\mu k^\nu \bar{\sigma}_{\mu\nu}\!\bigg)
      |1^a] ,
\end{align} \label{eq:Lorentz1to2Spinor}%
\ese
where $U_{12} \in {\rm SU}(2)$ is a little-group transformation
that depends on the specifics of the massive-spinor realization.
The duality properties of the spinorial generators~\eqref{eq:LorentzGeneratorSpinor}
allow us to easily rewrite the above exponents as
\be
   \frac{i}{m^2} p_1^\mu k^\nu \sigma_{\mu\nu,\alpha}^{~~~~\;\beta}
    = k \cdot a_{\alpha}^{~\beta} , \qquad
   \frac{i}{m^2} p_1^\mu k^\nu \bar{\sigma}^{~~~\dot{\alpha}}_{\mu\nu,~\dot{\beta}}
    =-k \cdot a^{\dot{\alpha}}_{~\,\dot{\beta}} ,
\label{eq:Tensor2Vector}
\ee
where we have defined chiral representations for the Pauli-Lubanski operators
\bse
\begin{align}
   a_{~~\alpha}^{\mu,~\,\beta} &
    = \frac{1}{2m^2} \epsilon^{\mu\nu\rho\sigma} p_{{\rm a}\:\!\nu}
      \sigma_{\rho\sigma,\alpha}^{~~~~\;\beta} , \\
   a^{\mu,\dot{\alpha}}_{~~~\,\dot{\beta}} &
    = \frac{1}{2m^2} \epsilon^{\mu\nu\rho\sigma} p_{{\rm a}\:\!\nu}
      \bar{\sigma}^{~~~\dot{\alpha}}_{\rho\sigma,~\dot{\beta}} .
\end{align} \label{eq:PauliLubanskiSpinor}%
\ese
(Note that the product $k \cdot a$ is insensitive
to the difference between $p_1$ and $p_{\rm a}=p_1+k/2$ in the above definitions,
so we could pick the latter for further convenience.)
Extension to the higher-spin states
represented by tensor products of massive spinors
is analogous to \eqn{eq:LorentzGenerator}, \eg
\beal
   (a^{\mu})_{\alpha_1\ldots\alpha_{2s}}^{~~~~~~~~\beta_1\ldots\beta_{2s}}
    = a_{~~\alpha_1}^{\mu,~~\beta_1}
      \delta_{\alpha_2}^{\beta_2}\!\ldots \delta_{\alpha_{2s}}^{\beta_{2s}} & \\
    + \ldots
    + \delta_{\alpha_1}^{\beta_1}\!\ldots \delta_{\alpha_{2s-1}}^{\beta_{2s-1}}
      a_{~~\alpha_{2s}}^{\mu,~~~\beta_{2s}} & ,
\label{eq:PauliLubanski2s}
\eeal
so we have
\beal
   \ket{2}^{\odot 2s}\!& = e^{k \cdot a}
      \big\{ U_{12} \ket{1} \big\}^{\!\odot 2s} , \qquad
   |2]^{\odot 2s}\!= e^{-k \cdot a} \big\{ U_{12} |1] \big\}^{\!\odot 2s} , \\
   \bra{2}^{\odot 2s}\!& = \big\{ U_{12} \bra{1} \big\}^{\!\odot 2s}
      e^{-k \cdot a} ,\:\,\quad
   [2|^{\odot 2s}\!= \big\{ U_{12} [1| \big\}^{\!\odot 2s} e^{k \cdot a} ,
\label{eq:Lorentz1to2SpinorPL}
\eeal
where the second line follows from the antisymmetry
of $\sigma^{\mu\nu}$ and $\bar{\sigma}^{\mu\nu}$ in the sense of
$ \epsilon^{\alpha\beta} \sigma_{~~~\beta}^{\mu\nu,~\gamma} \epsilon_{\gamma\delta} = -\sigma_{~~~\delta}^{\mu\nu,~\alpha} $.

Let us now inspect the spin dependence
of the three-point amplitudes.
In \rcite{Guevara:2018wpp}
we used their representation~\eqref{eq:GravityMatter3ptExp} for that.
In terms of the same Pauli-Lubanski operators~\eqref{eq:PauliLubanskiSpinor}, 
they can be rewritten in a simpler form:
\bse
\begin{align}
\label{eq:GravityMatter3ptPlus2}
   {\cal M}_3^{(s)}(k^+) & = -\frac{\kappa x^2}{2m^{2s-2}}
      [2|^{\odot 2s} e^{-2k \cdot a} |1]^{\odot 2s} , \\*
\label{eq:GravityMatter3ptMinus2}
   {\cal M}_3^{(s)}(k^-) & = -\frac{\kappa x^{-2}}{2m^{2s-2}}
      \bra{2}^{\odot 2s} e^{2k \cdot a}  \ket{1}^{\odot 2s} ,
\end{align} \label{eq:GravityMatter3pt2}%
\ese
which can be derived from \eqn{eq:GravityMatter3pt} using the identities
\bse
\begin{align}
   [1^a k] & = x \braket{1^a k} , \qquad
   [2^b k] = x \braket{2^b k} , \\
   [1^a 2^b] & = - \braket{1^a 2^b} + \frac{x}{m} \braket{1^a k} \braket{k\;\!2^b} .
\end{align}
\ese
The apparent spin dependence in the amplitude formulae above
is of the form $e^{\mp 2k \cdot a}$,
whereas there seems to be no such dependence
in the original formulae~\eqref{eq:GravityMatter3pt}
from \rcite{Arkani-Hamed:2017jhn}.
This apparent contradiction is resolved by taking into account the transformations~\eqref{eq:Lorentz1to2Spinor}:
the true angular-momentum dependence inherent
to the minimal-coupling amplitudes is independent of the spinorial basis.
(Indeed, it must also match the covariant formula~\eqref{eq:Arg}.)
For example, the plus-helicity amplitude~\eqref{eq:GravityMatter3ptPlus} involves $\braket{12}^{\odot 2s}$,
which in the chiral representation is simply
\be
   \braket{21}^{\odot 2s}
    = \big\{ U_{12} \bra{1} \big\}^{\!\odot 2s} e^{-k \cdot a} \ket{1}^{\odot 2s} ,
\ee
whereas in the antichiral representation it is
\be
   [2|^{\odot 2s} e^{-2k \cdot a} |1]^{\odot 2s}
    = \big\{ U_{12} [1| \big\}^{\!\odot 2s} e^{-k \cdot a} |1]^{\odot 2s} .
\ee
As pointed out in the Appendix, 
it is now natural to strip the spin-states to cleanly obtain the spin dependence.
Alternatively, in the classical (and arbitrary-spin) limit
we should treat the operator in-between as a C-number.
In that case, both expressions above become unambiguously
\be
   \lim_{s \to \infty} m^{2s} (U_{12})^{\odot 2s} e^{-k \cdot a} .
\ee
The factor of $m^{2s}$ cancels in the actual amplitudes:
\bse
\begin{align}
\label{eq:GravityMatter3ptPlus3}
   {\cal M}_3^{(\infty)}(k^+) & \approx -\frac{\kappa}{2} m^2 x^2
      e^{-k \cdot a}\!\lim_{s \to \infty} (U_{12})^{\odot 2s} , \\*
\label{eq:GravityMatter3ptMinus3}
   {\cal M}_3^{(\infty)}(k^-) & \approx -\frac{\kappa}{2} m^2 x^{-2}
      e^{k \cdot a}\!\lim_{s \to \infty} (U_{12})^{\odot 2s} .
\end{align} \label{eq:GravityMatter3pt3}%
\ese
The remaining unitary factor of $(U_{12})^{\odot 2s}$ parametrizes an arbitrary little-group transformation
that corresponds to the choice of the spin quantization axis~\eqref{eq:UnitSpinVector}.
As such, it is inherently quantum-mechanical
and therefore should be removed in the classical limit.
Indeed, it also appears in the simple product of polarization tensors
\begin{align}
   \lim_{s \to \infty} \varepsilon_2 \cdot \varepsilon_1
    = \lim_{s \to \infty} \frac{1}{m^{2s}}
      \braket{21}^{\odot s} \odot & [21]^{\odot s} \nn \\
\label{eq:SpinTransitionNorm}
    = \lim_{s \to \infty} \frac{1}{m^{2s}}
      \big\{ U_{12} \bra{1} \big\}^{\!\odot s} e^{-k \cdot a} &
      \ket{1}^{\odot s} \\ \times
      \big\{ U_{12} [1| \big\}^{\!\odot s} e^{k \cdot a} & |1]^{\odot s}
    = \lim_{s \to \infty} (U_{12})^{\odot 2s} , \nn
\end{align}
where $a$ is defined by \eqn{eq:PauliLubanski2s} but with half as many slots.
So we interpret the factor of $(U_{12})^{\odot 2s}$ as the state normalization
in accord with the notion of GEV of \rcite{Guevara:2018wpp}.

\section{Four-point amplitude
\label{sec:4pt}}

We are now ready to compute the four-point amplitude
that contains the complete information about
classical 1PM scattering of two spinning black holes,
with masses~$m_{\rm a}$ and~$m_{\rm b}$.
In the rigorous framework of Kosower, Maybee and O'Connell~\cite{Kosower:2018adc}
for computing the momentum deflection in the spinless case,
the tree-level contribution to the impulse expectation value is
\begin{align}\!\!
   \Delta p_{\rm a}^\mu &
    = \lim_{\hbar \to 0}
      \int\!\hat{d}^4p_1 \hat{\delta}_+(p_1^2-m_{\rm a}^2)
      \int\!\hat{d}^4p_2 \hat{\delta}_+(p_2^2-m_{\rm a}^2) \nn \\ & \quad~\times\!
      \int\!\hat{d}^4p_3 \hat{\delta}_+(p_3^2-m_{\rm b}^2)
      \int\!\hat{d}^4p_4 \hat{\delta}_+(p_4^2-m_{\rm b}^2) \\ & \quad~\times
      \hat{\delta}^4(p_1\!+p_3\!-p_2\!-p_4)
      \psi_{\rm a}(p_1) \psi_{\rm a}^*(p_2)
      \psi_{\rm b}(p_3) \psi_{\rm b}^*(p_4) \nn \\ & \quad~\times
      k^\mu e^{-ik\cdot b/\hbar} i{\cal M}_4(p_1,-p_2,p_3,-p_4) , \nn
\end{align}
where $k=p_2-p_1$ and
the wavefunctions~$\psi_{\rm a,b}$ describe quantum-mechanical wave packets
with momenta well approximated by the classical momenta $p_{\rm a,b}$.
It then leads to a schematic formula
\be
   \Delta p_{\rm a}^\mu
    = \bigg\langle\!\!\!\bigg\langle\!\int\!\hat{d}^4k\;\!
      \hat{\delta}(2p_{\rm a}\!\cdot k) \hat{\delta}(2p_{\rm b}\!\cdot k)
      k^\mu e^{-ik \cdot b/\hbar} i{\cal M}_4(k) \bigg\rangle\!\!\!\bigg\rangle .
\label{eq:ImpulseKMO}
\ee
Here the angle-bracket notation of \rcite{Kosower:2018adc}
involves a careful analysis of suitable wavefunctions~$\psi_{\rm a,b}$
and powers of~$\hbar$, and roughly amounts to
setting the momenta to their classical values as follows
\be
   k^\mu = \hbar \bar{k}^\mu \to 0 , \qquad
   p_1^\mu, p_2^\mu \to m_{\rm a} u_{\rm a}^\mu , \qquad
   p_3^\mu, p_4^\mu \to m_{\rm b} u_{\rm b}^\mu .
\label{eq:Momenta2Classical}
\ee

\begin{figure}
\includegraphics[width=0.4\textwidth]{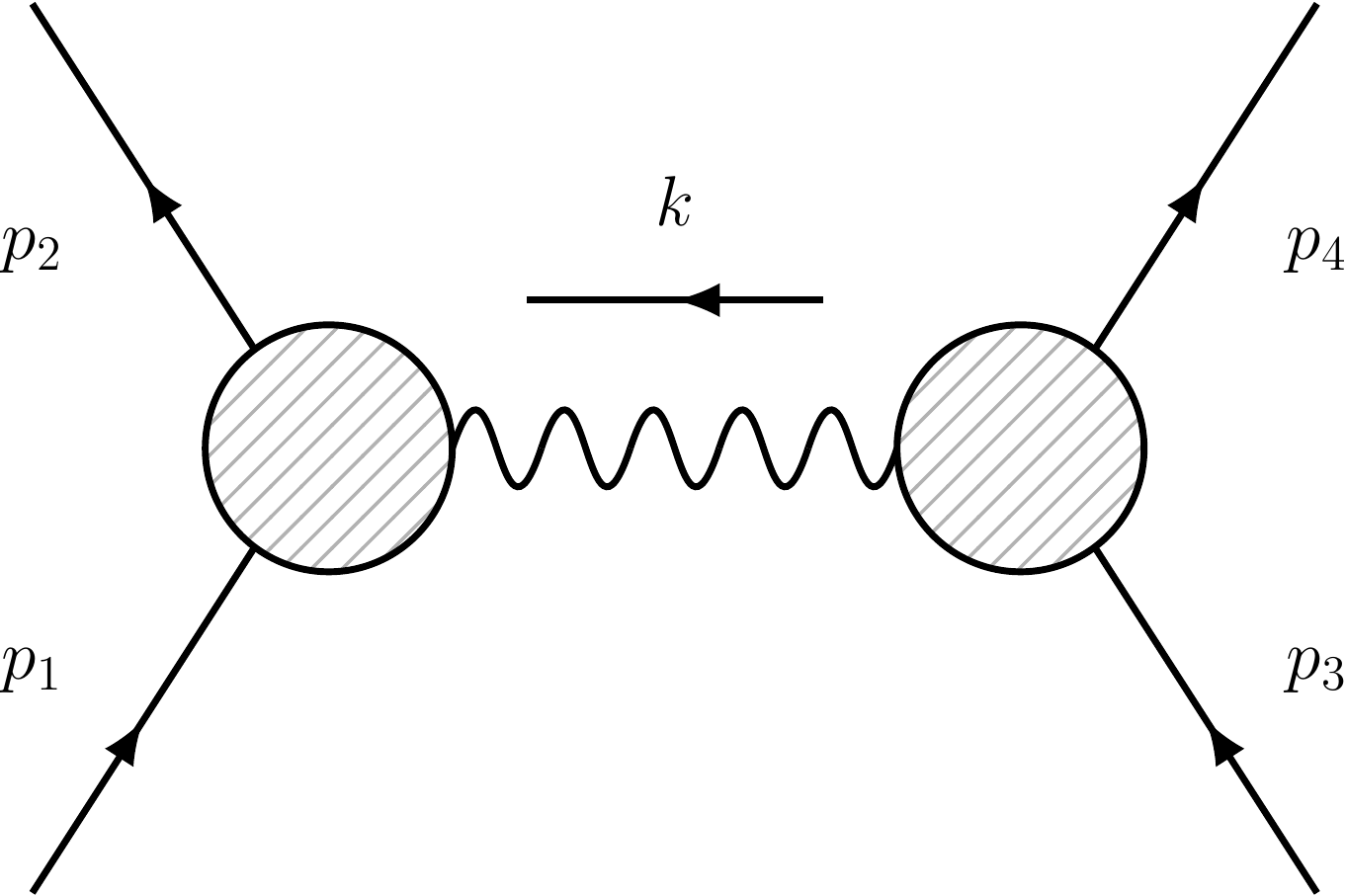}
\caption{\label{fig:4pt} Four-point amplitude for elastic scattering
of two distinct massive particles}
\end{figure}

First of all, we note that
in the quantum-mechanical setting of \rcite{Kosower:2018adc}
both $p_1$ and $p_2$ are associated
with the momentum of the first incoming black hole.
This is consistent with the equitable identification
\be
   p_{\rm a} = (p_1+p_2)/2 , \qquad
   p_{\rm b} = (p_3+p_4)/2 ,
\ee
that we will follow.
Moreover, the classical limit~\eqref{eq:Momenta2Classical}
prescribes inspecting soft-graviton exchange in the $t=k^2$ channel,
in which the graviton's momentum is taken to zero uniformly.
Here, however, we are going to adhere to an alternative strategy
of the HCL~\cite{Guevara:2017csg}:
we compute the residue of the scattering amplitude on the pole at $t=0$
on finite complex kinematics and analytically continue the result
to real kinematics at a later stage.
As shown in \fig{fig:4pt},
the four-point amplitude then conveniently factorizes into two three-point ones:
\begin{align}
\label{eq:SpinDeflectionAmplitude1}
 & {\cal M}_4^{(s_{\rm a},s_{\rm b})}(p_1,\!-p_2,p_3,\!-p_4) \\ &
    = \frac{\!-1}{t}\!\sum_{\pm}\!{\cal M}_3^{(s_{\rm a})}(p_1,\!-p_2,k^\pm)
      {\cal M}_3^{(s_{\rm b})}(p_3,\!-p_4,\!-k^\mp) + {\cal O}(t^0) \nn \\ &
    = \frac{-(\kappa/2)^2}{m_{\rm a}^{2s_{\rm a}-2} m_{\rm b}^{2s_{\rm b}-2} t}
      \Big( x_{\rm a}^2 x_{\rm b}^{-2}
            \braket{21}^{\odot 2s_{\rm a}} [43]^{\odot 2s_{\rm b}} \nn \\ &
            \qquad \qquad \qquad~\;\,\quad
          + x_{\rm a}^{-2} x_{\rm b}^2
            [21]^{\odot 2s_{\rm a}} \braket{43}^{\odot 2s_{\rm b}}
      \Big) + {\cal O}(t^0) , \nn
\end{align}
of which we now have complete understanding.

The helicity factors
enter the above amplitude in simple combinations evaluated on the pole kinematics as
\be
   x_{\rm a}/x_{\rm b} = \gamma(1-v) , \qquad
   x_{\rm b}/x_{\rm a} = \gamma(1+v) ,
\label{eq:MomentumDeflectionVariables}
\ee
where we have introduced the following interchangeable variables
that describe the total energy of the black-hole scattering process:
\be
   \gamma = \frac{1}{\sqrt{1-v^2}}
    = \frac{p_{\rm a}\!\cdot p_{\rm b}}{m_{\rm a} m_{\rm b}}
    = u_{\rm a}\!\cdot u_{\rm b} .
\label{eq:gammaFactor}
\ee
Evaluating the spin-dependent terms using \eqn{eq:Lorentz1to2SpinorPL}
and taking into account the direction of $k^\mu$, we get
\begin{align}
   {\cal M}_4^{(s_{\rm a},s_{\rm b})}
   = & \frac{-(\kappa/2)^2 \gamma^2 }
            { m_{\rm a}^{2s_{\rm a}-2} m_{\rm b}^{2s_{\rm b}-2} t } \nn \\ \times
      \Big( (1-v)^2
          & \big\{ U_{12} \bra{1} \big\}^{\!\odot 2s_{\rm a}}\;\!\!
            \exp(-k \cdot a_{\rm a})
            \ket{1}^{\odot 2s_{\rm a}} \nn \\ \times
          & \big\{ U_{34} [3| \big\}^{\!\odot 2s_{\rm b}}
            \exp(-k \cdot a_{\rm b})
            |3]^{\odot 2s_{\rm b}}
\label{eq:SpinDeflectionAmplitude2} \\
          + (1+v)^2
          & \big\{ U_{12} [1| \big\}^{\!\odot 2s_{\rm a}}
            \exp(k \cdot a_{\rm a})
            |1]^{\odot 2s_{\rm a}} \nn \\ \times
          & \big\{ U_{34} \bra{3} \big\}^{\!\odot 2s_{\rm b}}\:\!\!
            \exp(k \cdot a_{\rm b})
            \ket{3}^{\odot 2s_{\rm b}}
      \Big) + {\cal O}(t^0) . \nn
\end{align}
It is straightforward to check that the same result is obtained
if we choose to Lorentz-transform the states symmetrically to their averages:
$p_1, p_2 \to p_{\rm a}$ and $p_3, p_4 \to p_{\rm b}$.

Before we take the classical limit,
we should note that the above contractions of the Pauli-Lubanski pseudovector
are parity-odd.
To obtain a parity even expression, we observe that on the pole kinematics $k^2=0$
the Levi-Civita spin contractions satisfy
\bse
\begin{align}
   i\epsilon_{\mu\nu\rho\sigma} p_{\rm a}^\mu p_{\rm b}^\nu k^\rho a_{\rm a}^\sigma
    & = m_{\rm a} m_{\rm b} \gamma v (k \cdot a_{\rm a}) , \\
   i\epsilon_{\mu\nu\rho\sigma} p_{\rm a}^\mu p_{\rm b}^\nu k^\rho a_{\rm b}^\sigma
    & = m_{\rm a} m_{\rm b} \gamma v (k \cdot a_{\rm b}) .
\end{align} \label{eq:spin1pole}%
\ese
These equalities can be derived by squaring the left-hand sides
and computing the resulting Gram determinants using that
$ k^2 = p_{\rm a} \cdot k = p_{\rm b} \cdot k
= p_{\rm a} \cdot a_{\rm a} = p_{\rm b} \cdot a_{\rm b} = 0 $.
Therefore, introducing a two-form constructed from two initial BH momenta
\be
   w^{\mu\nu} = \frac{2p_{\rm a}^{[\mu} p_{\rm b}^{\nu]}}
                     {m_{\rm a} m_{\rm b} \gamma v} , \qquad
   [w*a]_\mu 
    = \frac{1}{2} \epsilon_{\mu\nu\alpha\beta} w^{\alpha\beta} a^\nu ,
\label{eq:wInForm}
\ee
and stripping the unitary transition factors
$U_{12}^{\odot 2s_{\rm a}}$ and $U_{34}^{\odot 2s_{\rm b}}$ via the GEV,
we obtain the classical limit of the scattering amplitude~\eqref{eq:SpinDeflectionAmplitude2}
as
\be
   \braket{{\cal M}_4(k)}
    =-\Big(\frac{\kappa}{2}\Big)^{\!2} \frac{m_{\rm a}^2 m_{\rm b}^2}{k^2}
      \gamma^2 \sum_\pm (1 \pm v)^2
      \exp[{\pm i} (k \cdot w*a_0)] ,
\label{eq:SpinDeflectionAmplitude3}
\ee
where $a_0^\mu = a_{\rm a}^\mu + a_{\rm b}^\mu$ is the total spin pseudovector.
Note that from now on we consider the above expression to be valid
for any values of transfer momentum momentum~$k$.

\begin{figure}
\includegraphics[width=0.2\textwidth]{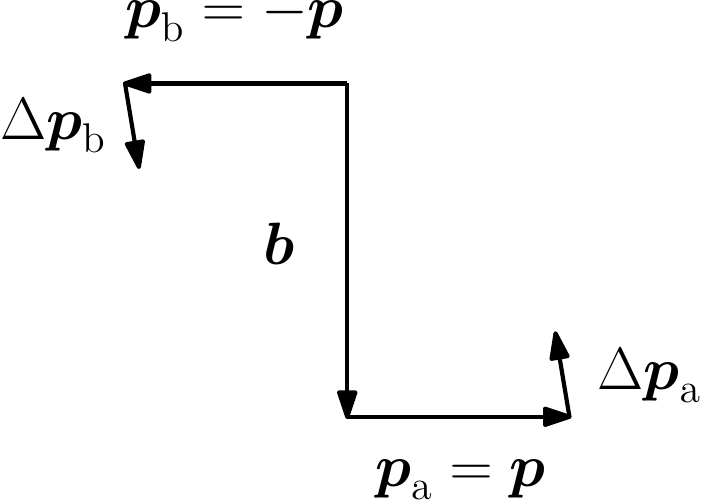}
\caption{\label{fig:COM} The BH three-momenta in the center-of-mass frame
and the impact parameter between them}
\end{figure}

As suggested by \eqn{eq:ImpulseKMO}
and the scattering-angle formula of \rcite{Bjerrum-Bohr:2018xdl},
in the classical picture we consider the transfer momentum~$k$
as a Fourier variable dual to the impact parameter $b$,
which is a spacelike vector orthogonal to both of initial momenta,
$b \cdot p_{\rm a} = b \cdot p_{\rm b} = 0$.
Therefore, we define the scattering function
\be
   \braket{{\cal M}_4(b)} = \int\!\hat{d}^4k\;\!
      \hat{\delta}(2p_{\rm a}\!\cdot k) \hat{\delta}(2p_{\rm b}\!\cdot k)
      e^{-ik \cdot b} \braket{{\cal M}_4(k)} .
\label{eq:k2b}
\ee
The above Fourier transform is easiest performed in the center-of-mass (COM) frame,
where $p_{\rm a} = (E_{\rm a}, \bs p)$ and $p_{\rm b} = (E_{\rm b}, -\bs p)$,
see \fig{fig:COM}.
In this frame
the eikonal integration measure~\cite{Bjerrum-Bohr:2018xdl} becomes explicitly
\beal
 & \int\!\hat{d}^4k\;\!\hat{\delta}(2p_{\rm a}\!\cdot k)
                       \hat{\delta}(2p_{\rm b}\!\cdot k) e^{-ik \cdot b} \\ &~\quad
    \overset{\text{COM frame}}{=}
      \frac{1}{4 (E_{\rm a}+E_{\rm b}) |\bs p|}
      \int\!\hat{d}^2 \bs k\,e^{i\bs k \cdot \bs b}
      \bigg|_{k^0 = \bs p \cdot \bs k = 0} .
\label{eq:Fourier2d}
\eeal
In other words, the integration is strictly spacelike
and restricted by $\bs p \cdot \bs k = 0$
to the two-dimensional subspace orthogonal to the initial momenta,
which is the same subspace where the impact parameter is defined.
Using $(E_{\rm a}+E_{\rm b}) |\bs p| = m_{\rm a} m_{\rm b} \gamma v$,
we compute
\begin{align}
\label{eq:SpinDeflectionAmplitudeFourierCOM}
 & \braket{{\cal M}_4(\bs b)}
    = \frac{\kappa^2 m_{\rm a} m_{\rm b} \gamma}{16 v}
      \sum_\pm (1 \pm v)^2 \\ & \quad\,\times\!
      \int\!\frac{d^2 \bs k}{(2\pi)^2 \bs k^2}
      \exp\!\big[ i \bs k \cdot \big(\bs b \mp [w*a_0]_{2d} \big)
            \big] \Big|_{\bs p \cdot \bs k = 0} \nn \\ &
    =-\frac{\kappa^2 m_{\rm a} m_{\rm b} \gamma}{32\pi v}
      \sum_\pm (1 \pm v)^2 \log\!\big|\bs b \mp [w*a_0]_{2d}\big| . \nn
\end{align}
Here by $[w*a_0]_{2d}$ we have denoted the appropriate spacelike projection
of the four-vector $w*a_0$.
However, recall that 
\be
   [w*a_0]^\mu
    = \frac{ \epsilon^{\mu\nu\rho\sigma}
             a_{0\:\!\nu} p_{\rm a\:\!\rho} p_{\rm b\:\! \sigma} }
           { m_{\rm a} m_{\rm b} \gamma v } ,
\label{eq:wInFormCOM}
\ee
\ie the vector $w*a_0$ is transverse to $p_{\rm a}$ and $p_{\rm b}$
and hence lies in the same plane as $\bs k$ and~$\bs b$.
Therefore, no information is lost in the two-dimensional projection above,
so we can safely uplift
the scattering function~\eqref{eq:SpinDeflectionAmplitudeFourierCOM}
to its Lorentz-invariant form
\be
   \braket{{\cal M}_4(b)}
    =-G m_{\rm a} m_{\rm b} \frac{\gamma}{v}
      \sum_{\pm} (1\pm v)^2
      \log\!\sqrt{-(b \mp w*a_0)^2} .
\label{eq:SpinDeflectionAmplitude}
\ee

\section{Linear and angular impulses
\label{sec:impulses}}

In this section
we relate the scattering function~\eqref{eq:SpinDeflectionAmplitude}
in the impact-parameter space
to the classical changes in linear and angular momentum of a BH
after gravitational scattering off another BH.
This problem was treated to all orders in spins at 1PM order
by one of the present authors~\cite{Vines:2017hyw},
producing the result (rewritten in the mostly-minus metric convention)
\begin{subequations} \begin{align}
   \Delta p_{\rm a}^\mu &
    = G m_{\rm a} m_{\rm b} \Re Z^\mu , \\
   \Delta a_{\rm a}^\mu &
    =-\frac{G m_{\rm b}}{m_{\rm a}}
      \big[ p_{\rm a}^\mu (a_{\rm a}\!\cdot\!\Re Z)
          + \epsilon^{\mu\nu\rho\sigma} (\Im Z_{\nu})
            p_{{\rm a}\:\!\rho} a_{{\rm a}\:\!\sigma} \big] ,\!
\end{align} \label{eq:Solution1PMgeneral}%
\end{subequations}
in terms of an auxiliary complex vector
\be
   Z^\mu
    = \frac{\gamma}{v} \sum_\pm \big(1 \pm v\big)^2
      [\eta^{\mu\nu} \mp i(*w)^{\mu\nu}]
      \frac{(b \mp w*a_0)_\nu}{(b \mp w*a_0)^2} .
\label{eq:Solution1PMgeneralZ2}
\ee
Now we can observe that differentiating the scattering function~\eqref{eq:SpinDeflectionAmplitude}
automatically produces its real and imaginary parts:
\bse
\begin{align}
   \frac{\partial~}{\partial b^\mu} \braket{{\cal M}_4(b)} &
    = -G m_{\rm a} m_{\rm b} \Re Z_\mu , \\
   \frac{\partial~}{\partial a^\mu} \braket{{\cal M}_4(b)} &
    = G m_{\rm a} m_{\rm b} \Im Z_\mu .
\end{align} \label{eq:Solution1PMgeneralZfromA}%
\ese
Using the known solution~\eqref{eq:Solution1PMgeneral}, we can identify
\begin{align}
\label{eq:Amp2Solution1PMgeneral}
   \Delta p_{\rm a}^\mu &
    =-\frac{\partial~}{\partial b_\mu} \braket{{\cal M}_4(b)} , \\
   \Delta a_{\rm a}^\mu &
    = \frac{1}{m_{\rm a}^2}
      \bigg[ p_{\rm a}^\mu a_{\rm a}^\nu \frac{\partial~}{\partial b^\nu}
          - \epsilon^{\mu\nu\rho\sigma} p_{{\rm a}\:\!\nu} a_{{\rm a}\:\!\rho}
            \frac{\partial~}{\partial a_{\rm a}^\sigma} \bigg]
      \braket{{\cal M}_4(b)} . \nn
\end{align}
At this point, we have merely matched
the derivatives of our scattering function~\eqref{eq:SpinDeflectionAmplitude}
to the known result~\eqref{eq:Solution1PMgeneral}.
Let us now promote this empirical matching to a derivation,
under the assumption that our approach is consistent
with that of \rcite{Kosower:2018adc}.

Indeed, the first line of \eqn{eq:Amp2Solution1PMgeneral} gives
precisely the impulse formula~\eqref{eq:ImpulseKMO} from \rcite{Kosower:2018adc}.
So let us focus on the second line.
In the concurrent preprint~\cite{Maybee:2019jus},
Maybee, O'Connell and one of the current authors have extended
the classical-limit approach of \rcite{Kosower:2018adc}
to include corrections in spin.
Their starting point for (the expectation value of)
the lowest-order angular impulse is
\begin{align}
\label{eq:AngularImpulseMOV}
   \Delta S_{\rm a}^\mu &
    = \bigg\langle\!\!\!\bigg\langle \int\!\hat{d}^4k\;\!
      \hat{\delta}(2p_{\rm a}\!\cdot k) \hat{\delta}(2p_{\rm b}\!\cdot k)
      e^{-ik \cdot b} \\ & \quad\:\:\times\!
      \bigg(\!{-}\frac{i}{m_{\rm a}^2}
             p_{\rm a}^\mu S_{\rm a}^\nu k_\nu {\cal M}_4(k)
           + \big[ S_{\rm a}^\mu, i{\cal M}_4(k) \big]
      \bigg) \bigg\rangle\!\!\!\bigg\rangle . \nn
\end{align}
Here the amplitude is considered to be a function of a one-particle spin vector
acting on the space of physical spin degrees of freedom,
\ie the little-group indices.
Therefore, we interpret it as the matrix element
\be
   (S_{\rm a}^\mu)_{\{a\}}^{~~~\{b\}}
    = (-1)^s \varepsilon_{{\rm a}\:\!\{a_1 \ldots a_{2s}\}}\!\cdot
             \Sigma^\mu\!\cdot
             \varepsilon_{\rm a}^{\{b_1 \ldots b_{2s}\}} ,
\ee
where the prefactor of $(-1)^s$ is due to the spacelike normalization
of the polarization vectors in \eqn{eq:PolTensorsNorm}.
The explicit form for such a spin vector at finite spin is given in \eqn{eq:PauliLubanskiLGOperator}.
It corresponds to the generator of the little-group transformations:
as an operator it satisfies the $\rm{so}(3)$ algebra
in the rest frame of $p_{\rm a}$.
This can also be stated covariantly as 
\be
   [S_{\rm a}^\mu,S_{\rm a}^\nu] 
    = \frac{i}{m_{\rm a}} \epsilon^{\mu\nu\rho\sigma}
      p_{{\rm a}\:\!\rho} S_{{\rm a}\:\!\sigma} .
\ee
As ${\cal M}_4$ is a function of $S_{\rm a}^{\mu}$, these $\rm{so}(3)$ rotations act as
\be
   [S_{\rm a}^\mu, {\cal M}_4]
    = \frac{i}{m_{\rm a}} \epsilon^{\mu\nu\rho\sigma}
      p_{{\rm a}\:\!\nu} S_{{\rm a}\:\!\rho}
      \frac{\partial {\cal M}_4}{\partial S_{\rm a}^\sigma} .
\ee
Therefore, we obtain the formula for the change in rescaled spin
\begin{align}
\label{eq:AngularImpulseMOV3}
   \Delta a_{\rm a}^\mu &
    = \frac{1}{m_{\rm a}^2} \bigg\langle\!\!\!\bigg\langle \int\!\hat{d}^4k\;\!
      \hat{\delta}(2p_{\rm a}\!\cdot k) \hat{\delta}(2p_{\rm b}\!\cdot k)
      e^{-ik \cdot b} \\ & \qquad~~\;\times\!
      \bigg[ p_{\rm a}^\mu a_{\rm a}^\nu (-ik_\nu)
           - \epsilon^{\mu\nu\rho\sigma} p_\nu a_{{\rm a}\:\!\rho}
      \frac{\partial~}{\partial a_{\rm a}^\sigma}
      \bigg] {\cal M}_4(k) \bigg\rangle\!\!\!\bigg\rangle , \nn
\end{align}
which maps directly to the second line of \eqn{eq:Amp2Solution1PMgeneral}.
Now that the impulse formulae~\eqref{eq:Amp2Solution1PMgeneral}
have meaning on their own,
we see that plugging the scattering function~\eqref{eq:SpinDeflectionAmplitude} gives
a novel derivation for the complete 1PM solution~\eqref{eq:Solution1PMgeneral}.

\section{Summary and Discussion
\label{sec:outro}}

In this paper we have obtained the dynamically complete solution to the (net) problem
of conservative spinning black-hole scattering at 1PM order as given in \rcite{Vines:2017hyw}, using minimal-coupling scattering amplitudes
with arbitrarily large quantum spin~\cite{Arkani-Hamed:2017jhn}.
We have rederived the spin-exponentiated structure of such three-point amplitudes
in four dimensions in a way that takes into account the Lorentz boost
between the incoming and outgoing momenta~\cite{Bautista:2019tdr}.
In \rcite{Guevara:2018wpp} this boost was overlooked but effectively restored
by the introduction of the generalized expectation value.
Here we have shown that considering this boost streamlines
the discussion of the spin exponentiation,
as well as allows for a cleaner connection
to the classical notion of spin in general relativity~\cite{Levi:2015msa}.

We have computed the change of the momentum and spin of the scattered black holes
at 1PM order using a four-point one-graviton-exchange amplitude,
which in the holomorphic classical limit~\cite{Guevara:2017csg}
is factorized into two three-point minimal-coupling amplitudes.
We have adopted the formulae of \rcite{Kosower:2018adc,Maybee:2019jus}
in a way that allowed us to extract the full spin dependence
of the linear and angular impulses of the black holes.
In this way, we obtained a complete match
to the known solution 1PM solution~\cite{Vines:2017hyw},
which allows for spins of the incoming black holes in arbitrary directions.
This is also a significant step forward from the simpler case
of the angular momenta aligned perpendicular to the scattering plane
considered in \rcite{Guevara:2018wpp}.
It is promising that our calculation displayed
a sufficiently uniform level of complexity all the way
between the starting point and the final result,
even despite the more complex nature of the quantum degrees of freedom.
This is thanks to the remarkable fact
that the spin multipoles of a black hole exponentiate~\cite{Hansen:1974zz,Vines:2017hyw},
which we could exploit and thus avoid explicit multipole expansions.

There are several interesting future directions.
One of the most relevant ones is the extension to higher loop, or PM, orders
\cite{Cachazo:2017jef,Bjerrum-Bohr:2018xdl,Cheung:2018wkq,Cristofoli:2019neg},
which may require to include radiative corrections
\cite{Kosower:2018adc,Bautista:2019tdr,Maybee:2019jus}
and finite-size effects \cite{Bini:2012gu,Levi:2014gsa,Chen:2019hac,Cai:2019npx}.
One could also attempt to consider higher curvature corrections,
such as the ones in \rcites{Brandhuber:2019qpg,Emond:2019crr}.
It is also interesting to explore the test-body 
limit~\cite{Vines:2017hyw,Vines:2018gqi}
to improve our understanding of the effective potential
in the sense of~\rcite{Antonelli:2019ytb}.
Furthermore, it may prove beneficial
to use the double-copy approach to quantum gravity amplitudes
\cite{Bern:2008qj,Bern:2010ue,Johansson:2014zca,Johansson:2015oia},
which has recently seen classical extensions 
\cite{Monteiro:2014cda,Goldberger:2016iau,Luna:2016hge},
and in this way apply it to the binary-inspiral problem
\cite{Goldberger:2017ogt,Goldberger:2017vcg,
Li:2018qap,Shen:2018ebu,Plefka:2018dpa,Bautista:2019tdr,Plefka:2019hmz}.
Additional insight may come from studying the same problem
in a supersymmetric setting~\cite{Caron-Huot:2018ape}.

In conclusion, our work, together with \rcites{Chung:2018kqs,Maybee:2019jus},
opens the way to higher-order calculations
for the spin effects in classical black-hole scattering
using the modern amplitude techniques in an on-shell and gauge-invariant framework.

\begin{acknowledgments}
We thank Fabi\'an Bautista, David Kosower, Ben Maybee and Donal O'Connell
for illuminating discussions and collaboration on related topics.
A.G. acknowledges support via Conicyt grant  21151647.
A.O. has received funding from the European Union's Horizon 2020 research and innovation programme under the Marie Sk\l{}odowska-Curie grant agreement 746138. Research at Perimeter
Institute is supported in part by the Government of Canada through the Department of
Innovation, Science and Economic Development Canada and by the Province of Ontario
through the Ministry of Economic Development, Job Creation and Trade.
\end{acknowledgments}

\appendix*
\section{Spin multipoles from boosts
\label{app:multipoles}}

Here we review the construction of \rcite{Bautista:2019tdr}
applied to the three-point amplitudes
and show how it simplifies in the spinor-helicity formalism.
Consider the three-point amplitude in the covariant form as given there: 
\be
   {\cal M}_3^{(s)} = {\cal M}_3^{(0)}\:\!
      \varepsilon_2 \cdot
      \exp\!\bigg(\!{-i}\frac{k_\mu \varepsilon_\nu \Sigma^{\mu\nu}}
                             {p_1\cdot\varepsilon}\bigg)
      \cdot \varepsilon_1 ,
\label{eq:Exp}
\ee
where $\varepsilon_1$ and $\varepsilon_2$
are spin-$s$ polarization tensors,
the generators $\Sigma^{\mu\nu}$ are given in \eqn{eq:LorentzGenerator},
and ${\cal M}_3^{(0)} = -\kappa (p_1\cdot\varepsilon)^2$
corresponds to the gravitational interaction of a scalar particle.
It was proposed that in order to extract classical multipoles
(forming representations of the little group
in the sense of \rcite{Levi:2015msa})
the spin states must be evaluated at the same momenta.
On the three-point kinematics,
the polarization states for $p_1$ and $p_2$ are related via
\be
   \varepsilon_2
    = \exp\!\bigg( \frac{i}{m^2} p_1^\mu k^\nu \Sigma_{\mu\nu}\!\bigg)
      \tilde{\varepsilon}_1 , \qquad
   \tilde{\varepsilon}_1 = U_{12}^{(s)} \varepsilon_1 ,
\label{eq:Boost1}
\ee
where $U_{12}^{(s)}$ is a tensor representation
of an ${\rm SO}(3)$ little-group transformation.
Note that in the rest frame of particle~1
the Lorentz transformation $p_1^\mu k^\nu \Sigma_{\mu\nu} = m k^i \Sigma^{0i}$
is nothing but the canonical choice for
the boost needed to shift $p_1$ to $p_1+k$.
One can show that the two exponents commute on the three-point kinematics, so
\begin{align}
   {\cal M}_3^{(s)}\!& = {\cal M}_3^{(0)}
      \tilde{\varepsilon}_1
      \exp\!\bigg(\!{-}\frac{i}{m^2} p_1^\mu k^\nu \Sigma_{\mu\nu}\!\bigg)
      \exp\!\bigg(\!{-i}\frac{k_\mu \varepsilon_\nu \Sigma^{\mu\nu}}
                             {p_1\cdot\varepsilon}\bigg)
      \varepsilon_1 \nn \\ &
    =  {\cal M}_3^{(0)}
      \tilde{\varepsilon}_1
      \exp\!\bigg(\!{-i}\frac{k_\mu \varepsilon_\nu \Sigma_\perp^{\mu\nu}}
                             {p_1\cdot\varepsilon}\bigg)
      \varepsilon_1 ,
\label{eq:Arg}
\end{align}
where we have defined
\be
   \Sigma_\perp^{\mu\nu}
    = \Sigma^{\mu\nu}
    + \frac{2}{m^2} p_1^{[\mu} \Sigma^{\nu]\rho} p_{1\:\!\rho} , \qquad
   p_{1\:\!\mu} \Sigma_\perp^{\mu\nu} = 0 ,
\label{eq:nonSSC2SSC}
\ee
as the operator that corresponds
to the transverse spin tensor~\eqref{eq:PauliLubanski2}.
Being a transverse tensor, it can be used to construct representations
of the little group.
The first such $j$ spin multipoles of \eg \rcite{Levi:2015msa}
are recovered by expanding the exponential to order~$j$
and stripping $\varepsilon_1$ and $\tilde{\varepsilon}_1$.
For finite spin $s$, it was observed
that this exponential truncates at order $2s$,
whereas \eqn{eq:Exp} truncates at order $s$ \cite{Bautista:2019tdr}. 

Let us now apply the spinor-helicity formalism to the above argument.
Picking for concreteness the negative helicity, it was shown in \cite{Bautista:2019tdr} that \eqn{eq:Exp} can be rewritten as our amplitude:
\be
   {\cal M}_3^{(s)} = \frac{{\cal M}_3^{(0)}}{m^{2s}} \bra{2}^{\odot 2s}
      \exp\!\bigg(\!{-i}\frac{k_\mu \varepsilon_\nu^- \sigma^{\mu\nu}}
                        {p_1\cdot\varepsilon^-}\bigg)
      \ket{1}^{\odot 2s} .
\ee
On the other hand, in \rcite{Guevara:2018wpp} we noted that in the chiral spinor-variable basis
self-duality of $\sigma^{\mu\nu}$ implies
\be
   {-i}\frac{k_\mu \varepsilon_\nu^- \sigma^{\mu\nu}}{p_1\cdot\varepsilon^-}
    = -2i\frac{k_\mu \varepsilon_\nu^- \sigma_\perp^{\mu\nu}}{p_1\cdot\varepsilon^-}
    = 2 k \cdot a ,
\label{eq:GOV}
\ee
where $a^\mu$ is given by \eqref{eq:PauliLubanski2s}
and $\sigma_\perp^{\mu\nu}$ is the transverse projection of $\sigma^{\mu\nu}$,
as in \eqn{eq:nonSSC2SSC}.
The crucial factor of two arises here because in the spinor variables we cannot
distinguish between the orbital or intrinsic pieces of the angular momentum.
Indeed, the $p_1 \to p_2$ boost considered in \eqn{eq:Boost1}
acts on the chiral basis as
\be\!
   \frac{i}{m^2} p_1^\mu k^\nu \sigma_{\mu\nu} = k \cdot a \quad \Rightarrow \quad
   \ket{2}^{\odot 2s}\!= e^{k \cdot a} \big\{ U_{12} \ket{1} \big\}^{\!\odot 2s}\!\!
\label{eq:Boost}
\ee
in accord with \eqns{eq:Tensor2Vector}{eq:Lorentz1to2SpinorPL} in the main text.
This boost compensates the factor of two in \eqn{eq:GOV}, so
\beal
   {\cal M}_3^{(s)} & = \frac{{\cal M}_3^{(0)}}{m^{2s}}
      \big\{ U_{12} \bra{1} \big\}^{\!\odot 2s}
      e^{-k \cdot a} e^{2k \cdot a} \ket{1}^{\odot 2s} \\ &
    = \frac{{\cal M}_3^{(0)}}{m^{2s}}
      \big\{ U_{12} \bra{1} \big\}^{\!\odot 2s} e^{k \cdot a} \ket{1}^{\odot 2s} .
\eeal
Now compare this to \eqn{eq:Arg}, where two distinct exponentials
combined into an exponential of a ${\rm SO}(3)$ rotation~\eqref{eq:nonSSC2SSC}.
We see that in the four-dimensional chiral spinor basis
it trivialized down to two exponentials, identical up to a numerical prefactor.

One might see an apparent contradiction in \eqn{eq:Boost}.
Namely, that the right-hand side involves the little-group
rotation $k \cdot a$ preserving $p_1$,
whereas the left-hand side corresponds to a boost $p_1 \to p_1+k$.
The reason that this is consistent is because \eqn{eq:Boost}
is a chirality-dependent statement.
In fact, the corresponding relation for the antichiral spinors
involves a sign flip:
\be
   \frac{i}{m^2} p_1^\mu k^\nu \bar{\sigma}_{\mu\nu} = -k \cdot a , \label{eq:noc2}
\ee
as given in \eqn{eq:Tensor2Vector}.
More concretely, consider the following relations
\bse
\begin{align}
   \big[e^{k \cdot a}\big]_\alpha^{\,\,\,\beta} p_{1\:\!\beta\dot{\beta}}
   \big[e^{-k \cdot a}\big]^{\dot{\beta}}_{\,\,\,\dot{\alpha}} &
    = p_{1\:\!\alpha\dot{\alpha}}, \\
   \big[e^{k \cdot a}\big]_\alpha^{\,\,\,\beta} p_{1\:\!\beta\dot{\beta}}
   \big[e^{k \cdot a}\big]^{\dot{\beta}}_{\,\,\,\dot{\alpha}} &
    = p_{2\:\!\alpha\dot{\alpha}},
\end{align}
\ese
where $p_{i\:\!\alpha\dot{\alpha}} = \ket{i^a}_\alpha [i_a|_{\dot{\alpha}}$
as usual.
The first relation is simply the statement that
the Pauli-Lubanski operator generates little-group rotations,
whereas the second relation shows that thanks to the sign flip
$k \cdot a$ can effectively act as a boost.
Therefore, \eqn{eq:Boost} and \eqn{eq:noc2} contain no real contradiction and reflect the `square root' nature of the spinor-helicity representation.

\bibliography{references}

\end{document}